# Lagrangian for RLC circuits using analogy with the classical mechanics concepts

**Albertus Hariwangsa Panuluh and Asan Damanik**

Department of Physics Education, Sanata Dharma University
Kampus III USD Paingan, Maguwoharjo, Sleman, Yogyakarta, Indonesia

Email: panuluh@usd.ac.id

**Abstract**. We study and formulate the Lagrangian for the LC, RC, RL, and RLC circuits by using the analogy concept with the mechanical problem in classical mechanics formulations. We found that the Lagrangian for the LC and RLC circuits are governed by two terms i. e. kinetic energy-like and potential energy-like terms. The Lagrangian for the RC circuit is only a contribution from the potential energy-like term and the Lagrangian for the RL circuit is only from the kinetic energy-like term.

## 1. Introduction

Lagrangian formalism is a powerful way to obtain the equation of motion of a physical system. The Lagrangian formalism is turned up to solve problems that are not simple by using Newtonian Mechanics [1]. In Newtonian mechanics, we usually formulate the mechanical problem (physical system) in the form of force or vector. Meanwhile, in Lagrangian mechanics, we approach the mechanical problem in the form of energy which is a scalar. If we know the Lagrangian of a physical system, then we can obtain the equation of motion of a physical system by using the Euler-Lagrange equation [2].

In the electrical subject, the RLC (Resistance-Inductance-Capacitance) circuit is commonly discussed in student physics textbooks as an example of the application of electronics principles to build an electronic device [3]. In the mathematical physics subject, the RLC circuit is commonly used to show the application of differential equation in the physical system [4]. The knowledge of RLC circuit is certainly of great physical interest both from experimental (applied) and theoretical sides. Recently, the RLC circuit and its physical properties to be an interesting research subject due to the development of mesoscopic physics and nanophysics that need a quantum mechanics because the mesoscopic system and nanophysics are not anymore a macroscopic system.

The analysis of RLC circuit as a mesoscopic system by using quantum mechanics based on Cardirola-Kanai Hamiltonian and quantum invariant method to solve the Schrödinger equation for the RLC circuit and to obtain the corresponding wave functions in term of a particular solution of Milne-Penney equation was reported by Pedrosa and Pinheiro [5]. The quantum mechanics treatment of the RLC circuit with discrete charge and semiclassical consideration can be found in Utreras-Diaz [6] that obtained the approximation of energy eigenvalues in term of a dimensionless parameter $\sqrt{L/C}\,e^2/h$ ($e$ is the electron charge, and $h$ is the Planck constant). In literature, we found most analysis of the RC, RL, LC, and RLC circuits, as a mesoscopic system or nanophysics, are commonly discussed and formulated by using the concept of quantum mechanics [7-10]. As we have already known that the energy operator in quantum mechanics is the Hamiltonian which is defined as the sum of energy kinetic and energy potential of the physical system. In classical mechanics, we also know that we can formulate the physical system in form of Lagrangian defined as the kinetic energy minus potential energy of the physical system. Thus, it is also an interesting problem to know the explicit form of the Lagrangian for RC, LC, RL, and RLC circuits and its phenomenological implications.

In this paper, we formulate the Lagrangian for LC, RC, RL, and RLC circuits by using the analogy with the classical mechanics formulation for a physical system. We also discuss the advantages of the lagrangian formulation of LC, RC, RL, and RLC circuits compare to the o electrical formulation as one can find in the most of the student textbooks. We organized this paper as follow: in section 2 we derive the Lagrangian for LC circuit; in section 3 we derive the Lagrangian formulation for the RC circuit; in section 4, we formulate the lagrangian for RL circuit; in section 5 we derive the Lagrangian for RLC circuit. Finally, in section 6 we write the conclusions.

## 2. Lagrangian for the LC Circuit

In this section, we formulate the Lagrangian for the LC circuit. We begin with by writing the differential equation of Kirchhoff's rule for LC circuit without source as follow

$$\frac{d^2 q}{dt^2} + \frac{q}{LC} = 0. \tag{1}$$

By analogy with the harmonic oscillator, we can put the angular frequency regarding L and C as follow

$$\omega = \sqrt{\frac{1}{LC}}. \tag{2}$$

As we know from the mathematical physics, the general solution of Eq. (1) is given by

$$q(t) = k_1 e^{i\omega t} + k_2 e^{-i\omega t} \tag{3}$$

where we can determine the constants $k_1$ and $k_2$ from the appropriate boundary values.

Mathematically, there is two possible initial conditions for the capacitor C. First, at time $t = 0$ the capacitor has zero charges: $q(0) = 0$ and at time $t = \infty$ the charge of the capacitor reach a maximum value that is $Q$ and this process is known as charging process. Second, at time $t = 0$ it is possible that the capacitor C has initial charge $Q_0$ and at time $t = \infty$ the capacitor has no charge which is known as discharging a capacitor. If we put the the initial condition $q(0) = 0$ into Eq. (3), then we have

$$q(0) = 0 = k_1 e^0 + k_2 e^0 \tag{4}$$

that implies

$$k_1 = -k_2 \tag{5}$$

By inserting Eq. (5) into Eq. (3), we obtain

$$q(t) = k_1 (e^{i\omega t} - e^{-i\omega t}). \tag{6}$$

If we use Euler's equation, then Eq. (6) reads

$$q(t) = 2i k_1 \sin(\omega t). \tag{7}$$

It is clear from Eq. (7) that the charge as a sine function. The solution of the differential equation of Eq. (1) with the initial condition $q(0) = 0$ is not relevant to the physical reality although it is possible from the mathematics point of view. Therefore we discard the charging process with the solution in the form of sine function.

If we put the initial condition $q(0) = Q_0 = k_1 + k_2$ for discharging process into Eq. (3), then we have

$$k_2 = Q_0 - k_1. \tag{8}$$

By inserting the Eq. (8) into the Eq. (3) and after doing a little algebra, the Eq. (3) become

$$q(t) = Q_0 e^{-i\omega t} + 2i k_1 \sin(\omega t). \tag{9}$$

The second term of Eq. (9) does not satisfy the initial condition that $t = \infty \rightarrow q = 0$, so we drop the second term of Eq. (9) and finally, we have the solution for discharging capacitor in the LC circuit as follow

$$q(t) = Q_0 e^{-i\omega t}. \tag{10}$$

Now, we are in the position to formulate the Lagrangian for LC circuit. To formulate the lagrangian of a physical system, we must know the potential energy and kinetic energy of the system. The potential energy for the LC circuit is given by

$$V(q) = \frac{1}{2C} q^2. \tag{11}$$

By analogy with the potential energy of harmonic oscillator, we can put the corresponding parameters as follow

$$m \leftrightarrow L, \quad x \leftrightarrow q, \quad k \leftrightarrow \frac{1}{C}. \tag{12}$$

By inserting Eq. (10) into Eq. (11), then we find the potential energy of the LC circuit as follow

$$V_{LC} = \frac{1}{2C} Q_0^2 e^{-2i\omega t}, \tag{13}$$

and the kinetic energy of oscillator harmonic system is

$$T_{ho} = \frac{1}{2} m \dot{x}^2. \tag{14}$$

By using the analogy in Eq. (12) and take derivative of Eq. (10), we have the kinetic energy of the LC circuit as follow

$$T_{LC} = -\frac{1}{2} L \omega^2 Q_0^2 e^{-2i\omega t}. \tag{15}$$

From Eqs. (13) and (15), we finally can write the Lagrangian for the LC circuit as follow

$$L_{LC} = -\frac{1}{2} L \omega^2 Q_0^2 e^{2i\omega t} - \frac{1}{2C} Q_0^2 e^{-2i\omega t}, \tag{16}$$

which is composed of two terms that can be put as kinetic energy-like and potential energy-like.

**3. Lagrangian for the RC Circuit**

In this section, we formulate the Lagrangian for the RC circuit. The differential equation of Kirchhoff's rule for RC circuit without a source is given by

$$R \frac{dq}{dt} + \frac{q}{C} = 0. \tag{17}$$

It has already known that there are also two processes that can be assigned to the RC circuit i.e. charging or discharging of the capacitor C.

By integrating Eq. (17), and using the values of $q = 0$ at time $t = 0$ and $q = q$ at the time $t = t$, we then have

$$q(t) = e^{-t/RC}. \tag{18}$$

From Eq. (18) it is apparent that the solution is unacceptable because if we put $t \to \infty$, then the charge becomes zero which is contrary to the physical process as a charging process. By integrating Eq. (18), and using the values of $q = Q_0$ at time $t = 0$ and $q = q$ at time $t = t$, then we obtain

$$q(t) = Q_0 e^{-t/RC}. \tag{19}$$

which is relevant to the physical process as previously stated. It is clear that there is no kinetic energy term for the RC circuit. The potential energy in the RC circuit as follows

$$V_{RC} = \frac{1}{2C} Q_0^2 e^{-2t/RC}. \tag{20}$$

Because there is no kinetic energy in the RC circuit, then the Lagrangian of RC circuit as follows

$$L_{RC} = \frac{1}{2C} Q_0^2 e^{-2t/RC}. \tag{21}$$

## 4. Lagrangian for the RL Circuit

The differential equation for the RL circuit is given by

$$L\frac{dI}{dt} + RI = 0. \tag{22}$$

The solution of the Eq. (22) read

$$I(t) = I_0 e^{-Rt/L}. \tag{23}$$

For the RL circuit, we can make an analogy with the equation of motion of a particle with the force is velocity dependent as follow

$$m\frac{dv}{dt} + kv = 0, \tag{24}$$

By putting $L$ correspond to mass $m$ and the resistance $R$ correspond to friction constant $k$, then we can put the current $I$ in Eq. (23) as a corresponding velocity in Eq. (24).

To formulate the kinetic energy of the RL circuit, we use the analogy concept by using the usual mechanical kinetic energy ½ $mv^2$ and electric current $I$ correspond with velocity $v$, and then we have

$$T_{RL} = \frac{1}{2} L I_0 e^{-2Rt/L} \tag{25}$$

as the kinetic energy term for RL circuit. For the RL circuit, there is no potential energy term. Thus, the lagrangian for the RL circuit read

$$L_{RL} = \frac{1}{2} L I_0 e^{-2Rt/L}. \tag{26}$$

## 5. Lagrangian for the RLC Circuit

The differential equation for the RLC circuit as follows

$$L\frac{d^2q}{dt^2} + R\frac{dq}{dt} + \frac{q}{C} = 0$$

or

$$\frac{d^2q}{dt^2} + \frac{R}{L}\frac{dq}{dt} + \frac{q}{LC} = 0 \tag{27}$$

By using analogy with a damped harmonic oscillator, we have

$$\gamma = \frac{R}{2L}, \qquad \omega_0 = \sqrt{\frac{1}{LC}} \tag{28}$$

where $\gamma$ and $\omega_0$ as damping factor and natural frequency respectively.

The general solution of Eq. (27) as follows

$$q(t) = C_1 e^{-(\gamma - \lambda)t} + C_2 e^{-(\gamma + \lambda)t} \tag{29}$$

where

$$\lambda = \sqrt{\gamma^2 - \omega_0^2}. \tag{30}$$

There are three possible cases for the parameter $\lambda$ i.e. overdamping, critical damping and underdamping.

### 5.1 Overdamping

Overdamping will occur when $\lambda > 0$ so the exponents in the first and second term in Eq. (29) are real. If we insert the discharging capacitor boundary condition into Eq. (29), then we have the solution as follow

$$q(t) = Q_0 e^{-(\gamma + \lambda)t}. \tag{31}$$

By using the same procedure to formulate the Lagrangian for LC circuit, we then have the Lagrangian for overdamping RLC circuit as follow

$$L_{RLCover} = \frac{1}{2}L(\gamma+\lambda)^2 Q_0^2 e^{-2(\gamma+\lambda)t} - \frac{Q_0^2 e^{-2(\gamma+\lambda)t}}{2C}. \qquad (32)$$

After doing a little algebra that is by expanding and simplifying $\lambda$ and $\gamma$, the Eq. (32) become

$$L_{RLCover} = \frac{1}{2}\left(\frac{R^2}{2L} - \frac{1}{C} + \frac{R}{2}\sqrt{\frac{R^2}{L^2} - \frac{4}{LC}}\right)Q_0^2 e^{-\left(\frac{R}{L}+\sqrt{\frac{R^2}{L^2}+\frac{4}{LC}}\right)t} - \frac{Q_0^2 e^{-\left(\frac{R}{L}+\sqrt{\frac{R^2}{L^2}+\frac{4}{LC}}\right)t}}{2C}. \qquad (33)$$

One can see that the Lagrangian for overdamping RLC circuit as shown in Eq. (33) is composed of two terms i. e. kinetic energy-like and potential energy-like.

*5.2 Critical Damping*
Critical damping will occur when $\lambda = 0$. As shown in [6], the Eq (29) becomes

$$q(t) = C_1 t e^{-\gamma t} + C_2 e^{-\gamma t}. \qquad (34)$$

By inserting the discharging capacitor boundary condition then Eq (34) becomes

$$q(t) = Q_0 e^{-\gamma t} \qquad (35)$$

and by using the same procedure as previously, we find the Lagrangian for the critical damping RLC circuit as follow

$$L_{RLCcrit} = -\frac{1}{8}\frac{R^2}{L}Q_0^2 e^{-Rt/L} - \frac{Q_0^2 e^{-Rt/L}}{2C}, \qquad (36)$$

which is composed of two terms i. e. energy kinetic like and potential energy-like.

*5.3 Underdamping*
The RLC circuit will undergo the underdamping when $\lambda < 0$, and the general solution for the case of underdamping RLC circuit can be read in [6]

$$q(t) = e^{-\gamma t}\left(A\cos(\omega_d t - \theta_0)\right) \qquad (37)$$

where

$$\beta = \sqrt{\omega_0^2 - \gamma^2} = \sqrt{\frac{1}{LC} - \frac{R^2}{4L^2}}. \qquad (38)$$

By inserting the discharging capacitor boundary condition and we set the initial phase angle $\theta_0$ by $0°$, then Eq (37) become

$$q(t) = e^{-\gamma t} Q_0 \cos(\beta t) \qquad (39)$$

and we obtain the lagrangian for the underdamping RLC circuit as follow

$$L_{RLCund} = \frac{1}{2}L\left(\frac{R}{2L}Q_0 e^{-\gamma t}\cos(\beta t) + \sqrt{\frac{1}{LC} - \frac{R^2}{4L^2}}Q_0 e^{-\gamma t}\sin(\beta t)\right)^2 - \frac{Q_0^2 e^{-Rt/2L}\cos^2(\beta t)}{2C} \qquad (40)$$

By inspecting Eq. (4), we can see that the first term is kinetic energy-like and the second term is potential energy-like.

It is clear from Eqs. (33), (36), and (40) that the Lagrangians for the RLC circuit are the sum of the kinetic energy-like and potential energy-like as well as one can find in classical mechanics. It can be stated that the Lagrangian terms of the RLC circuit independent of the value of $\lambda$ qualitatively.

**6. Conclusions**
We have studied and evaluated the Lagrangian for LC, RC, LR, and RLC circuits by using the analogy with the classical mechanics concept to formulate the lagrangian of a physical system. We find that the lagrangian for LC and RLC circuits are composed of terms that can be assigned as kinetic energy and potential energy terms in corresponding with the Lagrangian of a physical system in classical mechanics. Meanwhile, the Lagrangian for the RC circuit is only from potential energy-like contribution and the and the lagrangian for the LR circuit is only kinetic energy-like contribution. We also find that the contribution terms to the Lagrangian of the LRC circuit independent of the value of $\lambda$ qualitatively.